\documentclass[aps,twocolumn,prd,showpacs,notitlepage,nofootinbib,floatfix,showkeys]{revtex4-1}

\usepackage{blindtext}
\usepackage{hyperref}
\usepackage{amsmath,amssymb}
\usepackage{float}
\usepackage{microtype}

\usepackage{graphicx}
\usepackage{bm}
\usepackage{latexsym}
\usepackage{epsfig}
\usepackage{psfrag}
\usepackage{color}
\usepackage[dvipsnames]{xcolor}
\usepackage{subfigure}

\def\f{\frac}
\def\tf{\tfrac}
\def\l{\left}
\def\r{\right}
\def\d{{\rm d}}
\def\Mpl{M_{_{\mathrm{Pl}}}}

\def\be{\begin{equation}}
\def\ee{\end{equation}}
\newcommand{\Gm}{\Gamma_{\phi}}    
\newcommand{\nn}{\nonumber}
\newcommand{\e}{\textit{e}}

\newcommand{\viz}{\textit{viz.~}}
\newcommand{\ie}{\textit{i.e.~}}
\newcommand{\cf}{\textit{cf.~}}
\allowdisplaybreaks[1]

\begin{document}
\title{Accounting for the time evolution of the equation\\ of state parameter 
during reheating}
\author{Pankaj Saha}
\email{E-mail: pankaj@physics.iitm.ac.in}
\affiliation{Department of Physics, Indian Institute of Technology Madras, 
Chennai 600036, India}
\author{Sampurn~Anand}
\email{Current address: Department of Physics, School of Basic and Applied 
Science, Central University of Tamil Nadu, Thiruvarur 610005, India. 
E-mail: sampurn@cutn.ac.in}
\affiliation{Department of Physics, Indian Institute of Technology Madras, 
Chennai 600036, India}
\author{L.~Sriramkumar}
\email{E-mail: sriram@physics.iitm.ac.in}
\affiliation{Department of Physics, Indian Institute of Technology Madras, 
Chennai 600036, India}
\begin{abstract}
One of the important parameters in cosmology is the parameter characterizing 
the equation of state (EOS) of the sources driving the cosmic expansion.
Epochs that are dominated by radiation, matter or scalar fields, whether they
are probed either directly or indirectly, can be characterised by a unique 
value of this parameter.
However, the EOS parameter during reheating~~a phase succeeding inflation 
which is supposed to rapidly defrost our Universe--remains to be understood 
satisfactorily.
In order to circumvent the complexity of defining an instantaneous EOS 
parameter during reheating, an effective parameter~$w_\mathrm{eff}$, 
which is an average of the EOS parameter over the duration of reheating, 
is usually considered. 
The value of~$w_\mathrm{eff}$ is often chosen arbitrarily to lie in the range 
$-1/3 \leq w_\mathrm{eff} \leq 1$.
In this work, we consider the time evolution of the EOS parameter during 
reheating and relate it to inflationary potentials $V(\phi)$ that behave 
as $\phi^p$ around the minimum, a proposal which can be applied to a wide 
class of inflationary models.
We find that, given the index~$p$, the effective EOS parameter~$w_{\rm eff}$
is determined uniquely.
We discuss the corresponding effects on the reheating temperature and its
implications. 
\end{abstract}
\maketitle

\section{Introduction}

In order to explain the current observable Universe, the conventional hot big 
bang model requires very fine tuned initial conditions during the radiation 
dominated epoch.
This difficulty can be overcome if we assume that the universe went through a 
brief phase of nearly exponential expansion--an epoch dubbed as inflation--in 
its very early stages~\cite{Guth:1980zm,Starobinsky:1980te,Linde:1981mu}.
Apart from explaining the observed extent of isotropy of the cosmic microwave 
background~(CMB)~\cite{Bennett:2003bz,Peiris:2003ff,Ade:2013zuv,Ade:2015tva},
inflation also provides a natural mechanism to generate the small anisotropies
superimposed on the nearly isotropic background~\cite{Hawking:1982cz,
Guth:1982ec,Starobinsky:1982ee,Bardeen:1983qw}.
It is these CMB anisotropies which act as the seeds for the eventual formation 
of the large scale structure in the Universe~\cite{Lyth:1998xn}. 

\par

But, due to the accelerated expansion, inflation makes the universe cold and 
dilute.
To be consistent with big bang nucleosynthesis (BBN), the Universe must consist
of radiation and matter in thermal equilibrium, when its temperature is around 
$10\, {\rm MeV}$~\cite{Kawasaki:1999na,Steigman:2007xt,Fields:2014uja}. 
Inflation is typically driven with the aid of scalar fields, often referred 
to as the inflaton.
At the termination of inflation, the energy from the inflaton is supposed to 
be transferred to the particles constituting the standard model through a 
process called reheating~\cite{Albrecht:1982mp,Abbott:1982hn,Dolgov:1982th,
Traschen:1990sw}.
During this phase of reheating, the inflaton is expected to rapidly decay
producing matter and radiation in equilibrium, thereby setting the stage 
for the conventional hot big bang evolution. 

\par

The original mechanism for reheating, suggested soon after the idea of 
inflation was proposed, was based on the perturbative decay of the
inflaton~\cite{Albrecht:1982mp,Abbott:1982hn,Dolgov:1982th,Traschen:1990sw}.
However, about a decade later, it was realized that the perturbative 
mechanism does not capture the complete picture, as the decay of the 
inflaton was found to be dominated by non-perturbative processes. 
Importantly, it was recognized that, immediately after the termination 
of inflation, the inflaton acts like a coherently oscillating condensate 
which leads to parametric resonance of the fields coupled to the
inflaton~\cite{Kofman:1994rk,Shtanov:1994ce,Son:1996uv,Kofman:1997yn,
Boyanovsky:1995ud}. 
In fact, the initial stage of reheating is referred to as preheating, to
distinguish it from the later stage of perturbative decay.

\par

\begin{figure}[!t]
\begin{center}
\includegraphics[height=2.50cm,width=8.50cm]{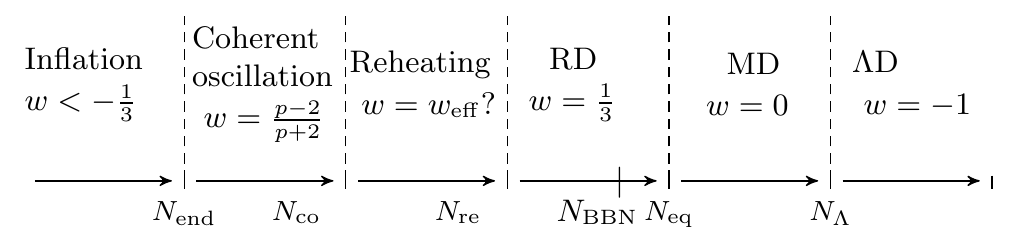}
\end{center}
\caption{A schematic timeline of cosmic evolution, with each epoch described 
by its respective EOS parameter.
The quantities $N_{\mathrm{end}}$, $N_{_{\mathrm{BBN}}}$, $N_{\mathrm{eq}}$ 
and $N_{\Lambda}$ refer to the \e-folds corresponding to the end of inflation, 
the epochs of BBN, radiation-matter equality and the beginning of 
$\Lambda$ domination, respectively.
Note that, however, $N_{\mathrm{co}}$ and $N_{\mathrm{re}}$ refer to the 
{\it duration}\/ of the phases of coherent oscillations and reheating.}
\label{fig:timeline}
\end{figure}

The details of the perturbative as well as the non-perturbative processes 
taking place during reheating can be non-trivial and will actually depend 
upon the various fields that are taken into account and the nature of their
interactions. 
Moreover, the lack of direct observables that can reveal the dynamics during 
this phase poses additional challenges towards understanding the mechanism of
reheating. 
In such a situation, as a first step, it would be convenient to characterise 
the phase through an equation of state (EOS) parameter $w$ which captures the
background evolution and, consequently, the dilution of the energy density 
of the fields involved, without going into the complexity of models and
interactions. 
After all, the different epochs of the Universe--\viz inflation, radiation
and matter domination as well as late time acceleration~---~are often simply
characterized in terms of the corresponding EOS parameter (in this 
context, see Fig.~\ref{fig:timeline}).
One widely adopted approach is to define an effective EOS 
parameter~$w_{\mathrm{eff}}$, which is an average of the
instantaneous EOS parameter during the period of
reheating~\cite{Martin:2010kz}. 
Although the averaging washes out the details of the microphysics over 
the intermediate stages, it allows us to conveniently characterize the 
reheating phase in terms of two other vital observables, \viz the 
duration of the phase and the reheating temperature. 
While such an approach may be adequate as a first step, needless to add, 
it is important to characterize and understand the dynamics of reheating
in further detail.

\par

As we mentioned, at the end of inflation,
the inflaton starts to oscillate about the minimum of the potential.
During the initial stages of this phase, most of the energy is stored 
in the coherently oscillating scalar field. 
It can be shown that the EOS parameter of a homogeneous condensate 
oscillating in a potential which has a minimum of the form $V(\phi)
\propto\phi^p$ is given by $w_\mathrm{co} =(p-2)/(p+2)$~\cite{Turner:1983he,
Mukhanov:2005sc,Podolsky:2005bw}. 
(For a brief discussion in this context for the case of 
$p=2$, see App.~\ref{appendix:appA}.)
However, in the process, the homogeneous condensate fragments leading 
to the growth of the inhomogeneities~\cite{Micha:2002ey,Micha:2004bv,
Allahverdi:2010xz,Amin:2014eta}. 
As a result, the EOS parameter differs from the above-mentioned form. 
The time when the EOS starts to change from its form during the 
period of coherent oscillations is referred to in the literature as the onset 
of the phase of backreaction~\cite{Figueroa:2016wxr,Maity:2018qhi}. 
The effects of fragmentation on the EOS can be studied using 
lattice simulations and one finds that the EOS parameter indeed eventually
approaches that of the radiation dominated phase (\ie $w\to 1/3$), as 
required~\cite{Lozanov:2016hid,Lozanov:2017hjm,Maity:2018qhi}. 
Evidently, the average $w_\mathrm{eff}$ during reheating will depend on the
time evolution of the EOS parameter from the end of coherent oscillations to 
the start of the radiation domination epoch.
Usually, the value of $w_\mathrm{eff}$ during this phase is either identified
to be the value $w_\mathrm{co}$ during the coherent oscillation phase or chosen
arbitrarily to lie in the range $-1/3 \leq w_\mathrm{eff} 
\leq 1$~\cite{Munoz:2014eqa}.

\par

\begin{figure}[!t]
\begin{center}
\includegraphics[width=8.50cm]{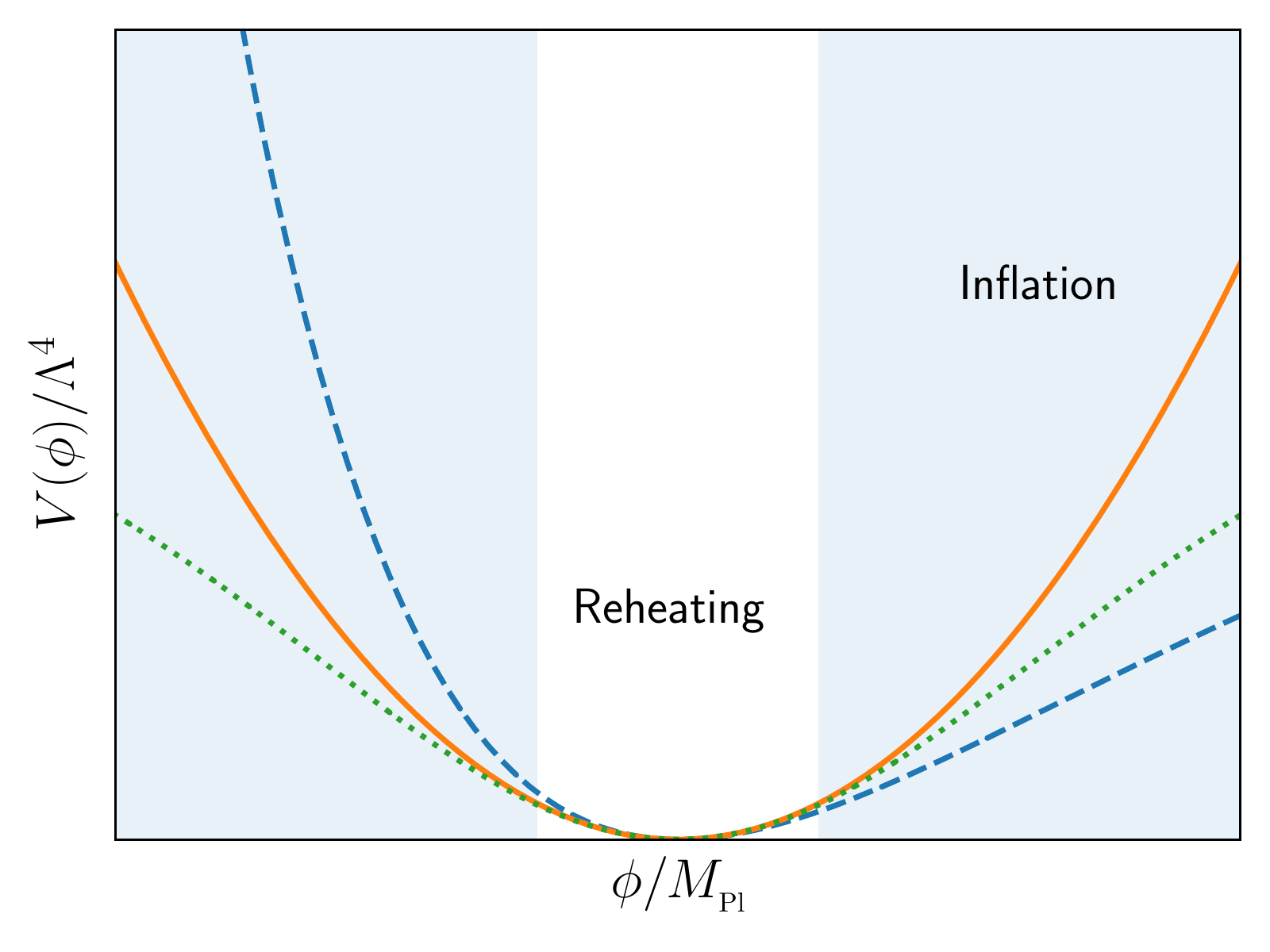}
\end{center}
\vskip -15pt
\caption{A schematic diagram illustrating the behavior of typical 
inflationary potentials of our interest around their minima.
We have plotted the potentials for the cases of the Starobinsky model 
(in blue), the so-called $\alpha$-attractor $T$-model (in green) and 
the quadratic potential (in orange).
Away from the minimum, at large field values, the presence of a plateau in 
the potentials (such as in the Starobinsky and $T$-models) ensure that the
inflationary predictions are consistent with the CMB observations. 
Around the minimum, the potentials behave as $V(\phi)\propto \phi^p$, 
which permits coherent oscillations during the initial stages of reheating.}\label{fig:schem_pot}
\end{figure}

In this work, we examine the time evolution of the EOS parameter and its 
average $w_\mathrm{eff}$ during reheating.
We consider the time evolution of the EOS from the end of the 
coherent oscillation stage until the onset of the radiation domination 
epoch. 
We argue that the presence of gradient and/or interaction energy of the 
inflaton leads to the deviation of the EOS parameter from its 
value~$w_\mathrm{co}$ during the period of coherent oscillations. 
Not surprisingly, we find that, even after the phase of coherent 
oscillations, the shape the inflationary potential near its minimum 
plays a role in the time evolution of the EOS.
We shall assume that, near their minima, the inflationary models of
our interest have the following form: $V(\phi) \propto \phi^p$. 
We should point out here that large field models which are {\it completely}\/ 
described by such power law potentials are already ruled out due to the 
constraints from the CMB data on the primary inflationary observables, 
\viz the scalar spectral index $n_{_{\mathrm{S}}}$ and tensor-to-scalar 
ratio~$r$~\cite{Komatsu2011wmap,Ade:2015lrj,Akrami:2018odb}. 
In contrast, potentials that contain a plateau, such as the original 
Starobinsky model, which lead to smaller values of~$r$ are favored by 
the CMB data. 
However, such potentials too can be expressed as $V(\phi)\propto \phi^p$ 
around the minima (in this context, see Fig.~\ref{fig:schem_pot} wherein
we have schematically illustrated the potentials for $p=2$). 
Motivated by results from lattice simulations, in order to 
capture the microphysics during reheating and, specifically, the turbulent
backreaction phase, we propose an evolving EOS parameter, which asymptotically
approaches its value during the radiation dominated epoch from its value at 
the end of the phase of coherent oscillations.
With such a time evolving EOS parameter, we establish a link between the 
value of $w_\mathrm{eff}$ and the inflationary potential parameters. 
This allows us to connect the reheating temperature uniquely to the 
inflationary parameters, while, importantly, accounting for the time
evolution of the EOS. 

\par

The remainder of the paper is structured as follows.
In Sec.~\ref{sec:review}, we shall provide a rapid overview of reheating and 
connect the parameters describing the phase with the observables in the CMB. 
In Sec.~\ref{sec:eos}, we shall first briefly highlight the motivations for
accounting for the time-dependence of the EOS during reheating. With the help of specific examples, we shall also present 
results from lattice simulations illustrating the time evolution of the EOS
parameter from the end of the coherent oscillation phase to the onset of 
the radiation dominated epoch.
We shall then go on to consider two types of parametrizations for the EOS 
parameter and arrive at the associated effective EOS parameter.
In Sec.~\ref{sec:im}, we shall apply these arguments to the so-called 
$\alpha$-attractor model of inflation and evaluate the corresponding
reheating temperatures for these models. 
We shall conclude with a brief summary of our results in Sec.~\ref{sec:conc}.


\section{Connecting the reheating phase with the CMB 
observables}\label{sec:review}

While the period of reheating is phenomenologically rich, as we mentioned, 
it is difficult to observationally constrain the dynamics due to the paucity 
of direct access to that epoch. 
Another difficulty arises due to the fact that by the time of BBN, all the 
particles associated with the standard model are expected to have been 
thermalized, thereby possibly hiding away the details of their production. 
Despite these limitations, one finds that reheating can still be constrained 
to a certain extent from the CMB and BBN observables. 
The upper bound on the inflationary energy scale, inferred from the 
constraints on the tensor-to-scalar ratio~$r$ (arrived at originally 
from the WMAP data~\cite{Komatsu2011wmap} and improved upon later by 
the Planck data~\cite{Ade:2015lrj,Akrami:2018odb}), is closer to the 
GUT scale of about $10^{16}\,{\mathrm{GeV}}$, whereas BBN requires a 
radiation dominated Universe at 
around $10\,{\mathrm{MeV}}$~\cite{Kawasaki:1999na,Steigman:2007xt,Fields:2014uja}. 
The inflationary observables are either well measured or have bounds on 
them, while the physics of BBN have been tested with great precision. 
Thus, there is a huge window in energy scales of several order of 
magnitudes which remains unconstrained by the cosmological data. 

\par

However, as has been pointed out in the literature, a connection can be 
made between the reheating phase and the CMB observables measured 
today~\cite{Martin:2014nya,Dai:2014jja,Munoz:2014eqa}. 
As it proves to be essential for our discussion later on, we shall quickly 
summarize the primary arguments in this section.
Recall that, during inflation, a scale of interest described by the comoving
wave number~$k$ leaves the Hubble radius at the time when $k = a_{k}\,H_{k}$.
Let this time correspond to, say, $N_k$, \e-folds {\it before the end of 
inflation}.\/
For instance, the Planck team choose their pivot scale to be $k = 0.05\,
{\mathrm{Mpc}^{-1}}$ and {\it assume}\/ $N_k\simeq 50$ for this
scale~\cite{Ade:2015lrj,Akrami:2018odb}. 
The physical wave number $k/a_k$ at the time when it exits the Hubble radius 
during inflation can be related to its corresponding value $k/a_0$ at the 
present time as follows: 
\begin{equation}
\frac{k}{a_k\, H_k} = \frac{k}{a_0\,H_k}\frac{a_0}{a_{\mathrm{re}}}\,
\frac{a_{\mathrm{re}}}{a_{\mathrm{co}}}\,
\frac{a_{\mathrm{co}}}{a_{\mathrm{end}}}\,
\frac{a_{\mathrm{end}}}{a_k},
\end{equation}
where $a_{\mathrm{end}}$, $a_{\mathrm{co}}$ and $a_{\mathrm{re}}$ are the values 
of the scale factor when inflation, the phase of coherent oscillations and 
reheating end, respectively.
Since ${\mathrm e}^{N_k} = a_{\mathrm{end}}/{a_k}$, ${\mathrm e}^{N_{\mathrm{co}}} 
= a_{\mathrm{co}}/a_{\mathrm{end}}$ and ${\mathrm e}^{N_{\mathrm{re}}} 
= a_{\mathrm{re}}/a_{\mathrm{co}}$, we can express the above equation as
\begin{equation}
N_k + N_{\mathrm{co}}+ N_{\mathrm{re}} 
+ \ln\left(\frac{a_0}{a_{\mathrm{re}}}\right)
+\ln\left(\frac{k}{a_0\, H_k}\right)  = 0,\label{eq:eqk1}
\end{equation}
where, evidently, $N_{\mathrm{co}}$ and $N_{\mathrm{re}}$ denote the 
{\it durations}\/ of the phase of coherent oscillations and the 
backreaction phase.  

\par

At the end of reheating, the Universe is supposed to be radiation dominated 
and if no significant entropy is released into the primordial plasma, we can 
relate the reheating temperature, say, $T_{\mathrm{re}}$, to the present CMB 
temperature, say, $T_0$, as follows (see, for example, Ref.~\cite{Munoz:2014eqa}):
\begin{equation}
\f{T_{\mathrm{re}}}{T_{0}}
=\l(\frac{43}{11\, g_{\mathrm{s}, \mathrm{re}}}\r)^{1 / 3}\, 
\frac{a_{0}}{a_{\mathrm{re}}},\label{eq:TreT02}
\end{equation}
where $g_{\mathrm{s},\mathrm{re}}$ denotes the effective number of relativistic
degrees of freedom that contribute to the entropy during reheating. 
We should mention that, to arrive at the above expression, we have expressed
the neutrino temperature in terms of the temperature~$T_0$ of the CMB using 
the relation $T_{\nu 0}=(4/11)^{1/3}\, T_{0}$. 
On using Eqs.~\eqref{eq:eqk1} and \eqref{eq:TreT02}, we can express the reheating
temperature as
\begin{equation}
T_{\mathrm{re}} =\l(\frac{43}{11\, g_{\mathrm{s}, \mathrm{re}}}\r)^{1 / 3}\,
\l(\frac{a_{0}\, T_{0}}{k}\r)\, H_k\, {\mathrm{e}}^{-N_{k}}\, 
{\mathrm{e}}^{-N_{\mathrm{co}}}\, {\mathrm{e}}^{-N_{\mathrm{re}}}.
\label{eq:Tre1}
\end{equation}

\par

Let us now assume that the backreaction phase succeeding the period of 
coherent oscillations is described by the time-dependent EOS parameter~$w(N)$.
In such a case, from the conservation of energy, the cosmic energy density 
during the phase can be expressed as
\begin{equation}
\rho(N)=\rho_{\mathrm{co}}\,
\exp\l\{-3\,\int_{0}^{N}\,\d N'\l[1+w\l(N'\r)\r]\r\},
\end{equation}
where $\rho_{\mathrm{co}}$ is the energy density at the end of the coherent 
oscillation phase.
On defining an averaged EOS parameter as
\be
w_{\mathrm{eff}}= \frac{1}{N_{\mathrm{re}}}\,
\int_0^{N_{\mathrm{re}}}\d N'\, w(N'),\label{eq:weff}                  
\ee
we can rewrite the above expression as
\be
\ln\left(\frac{\rho_{\mathrm{co}}}{\rho_{\mathrm{re}}}\right)
= 3\, (1+w_{\mathrm{eff}})\,N_{\mathrm{re}},\label{eq:rhoendNre}
\ee
where $N_{\mathrm{re}}$ denotes the number of \e-folds {\it during}\/ the
backreaction phase counted from the end of the period of coherent 
oscillations.

\par

If we now assume that, at the end of reheating, the dominant component of 
energy is radiation, then we can express the energy density of radiation 
in terms of~$T_{\mathrm{re}}$ as 
\begin{equation}
\rho_{\mathrm{re}} \equiv \rho_{\gamma}(T_{\mathrm{re}}) 
=\frac{\pi^{2}\, g_{\mathrm{re}}}{30}\, T_{\mathrm{re}}^{4},\label{eq:rhog}
\end{equation}
where $g_{\mathrm{re}}$ is the number of effective relativistic degrees of 
freedom at the end of reheating. 
In such a case, upon using Eqs.~\eqref{eq:rhoendNre} and~\eqref{eq:rhog},
we can readily express $T_{\mathrm{re}}$ as
\begin{equation}
T_{\mathrm{re}}
= \l(\f{30\,\rho_{\mathrm{co}}}{g_{\mathrm{re}}\,\pi^2}\r)^{1/4}\,
\mathrm{exp} \l[-\f{3}{4}(1+w_{\mathrm{eff}})\,N_{\mathrm{re}}\r].
\label{eq:Tre2}
\end{equation}
From Eqs.~\eqref{eq:Tre1} and~\eqref{eq:Tre2}, we can then arrive at the 
following expression for the duration $N_{\mathrm{re}}$ of the phase of 
reheating:
\begin{eqnarray}
N_{\mathrm{re}}
&=&\frac{4}{3\, w_{\mathrm{eff}}-1}
\Biggl[ N_{k} + N_{\mathrm{co}} + \ln \l(\frac{k}{a_{0} T_{0}}\r)\nn\\
& &+\,\f{1}{4}\,\mathrm{ln}\l(\f{30}{\pi^{2}\, g_{\mathrm{re}}}\r)
+ \frac{1}{3}\, \mathrm{ln} \l(\frac{11\, g_{\mathrm{s},{\mathrm{re}}}}{43}\r)\nn\\
& &-\, \ln \l(\f{H_{k}}{\rho_{\mathrm{end}}^{1/4}}\r) 
- \frac{1}{4}\,\ln \l(\f{\rho_{\mathrm{end}}}{\rho_{\mathrm{co}}}\r)\Biggr],
\label{eq:Nre}
\end{eqnarray}
where $\rho_{\mathrm{end}}$ is the energy density of the inflaton at the end 
of inflation.
Since the EOS parameter during the phase of coherent oscillations is 
$w_{\mathrm{co}} = (p-2)/(p+2)$, which is obviously a constant for 
given value of~$p$, we can express $\rho_{\mathrm{co}}$ in terms of 
$\rho_{\mathrm{end}}$ as 
\begin{equation}
\ln \l(\f{\rho_{\mathrm{end}}}{\rho_{\mathrm{co}}}\r) 
= 3\, \l(1 + \f{p-2}{p+2}\r)\,N_{\mathrm{co}}.
\end{equation}
Therefore, the duration of reheating $N_{\mathrm{re}}$ can finally be
expressed as
\begin{eqnarray}
N_{\mathrm{re}} &=&\frac{4}{3\, w_{\mathrm{eff}}-1}
\Biggl[N_{k} + \f{(4-p)}{2\,(p+2)}\,N_{\mathrm{co}}
+\ln \l(\frac{k}{a_{0}\, T_{0}}\r)\nn\\
& &+\,\f{1}{4}\ln\l(\f{30}{\pi^{2}\, g_{\mathrm{re}}}\r) 
+ \f{1}{3}\, \ln \l(\frac{11\, g_{{\mathrm{s},\mathrm{re}}}}{43}\r)
- \ln \l(\f{H_{k}}{\rho_{\mathrm{end}}^{1/4}}\r)\Biggr].\nn\\
\label{eq:Nre-f}
\end{eqnarray}

\par

During inflation, the energy density of the inflaton can be expressed in 
terms the potential $V(\phi)$ and the first slow roll parameter $\epsilon
=-\dot{H}/H^2$ as
\begin{equation}
\rho = V\,\l(1 +\f{\epsilon}{3-\epsilon}\r).
\end{equation}
Since inflation ends when $\epsilon = 1$, we have  $\rho_{\mathrm{end}} 
= (3/2)\,V_{\mathrm{end}}$, where $V_{\mathrm{end}}$ denotes the potential at 
$\phi_{\mathrm{end}}$, \viz the value of the scalar field at which inflation
is terminated.
Given the potential $V(\phi)$, the value of $\phi_{\mathrm{end}}$ can be readily
determined using the condition $\epsilon \simeq (M_{_{\mathrm{Pl}}}^2/2)\,
(V_{\phi}/V)^2 = 1$, where the subscript on the potential denotes the derivative
with respect to the scalar field.
Also, working in the slow roll approximation, we can calculate the value of the 
scalar field at~$N_k$.
This, in turn, can be utilized to express~$N_k$ in terms of the inflationary 
observables, \viz the scalar spectral index~$n_{_{\mathrm{S}}}$ and 
tensor-to-scalar ratio~$r$. 
Therefore, the bounds on the inflationary parameters from the CMB will lead 
to the corresponding constraints on the reheating parameters as well (in this 
context, see Refs.~\cite{Dai:2014jja,Munoz:2014eqa,Cook:2015vqa,Drewes:2017fmn,
Gong:2015qha,Maity:2018dgy,Maity:2018exj,Allahverdi:2018iod,Maity:2019ltu,
DiMarco:2019czi}). 
However, note that the quantity $N_{\mathrm{co}}$ depends on the details of the
inflationary model under investigation and, importantly, on the coupling of the 
inflaton to the other fields.
  

\section{Time-dependent EOS}\label{sec:eos}

As discussed earlier, the EOS parameter for the homogeneous condensate, 
oscillating about the minimum of a potential behaving as $V(\phi)\propto 
\phi^p$, is given by $w_{\mathrm{co}}
= (p-2)/(p+2)$~\cite{Mukhanov:2005sc,Turner:1983he}. 
However, due to the growth of inhomogeneities, the EOS parameter can be 
expected to differ from the above value during the backreaction phase. 
We can study the resulting variation in the EOS parameter by considering
virialization of the inhomogeneous system in equilibrium. 

\par

Consider a situation wherein the inflaton $\phi$ decays into daughter fields 
collectively represented as~$\mathcal{F}$ through the interaction potential
$V_{\mathrm{I}}(\phi,\mathcal{F})$.
For a potential that behaves as $V(\phi)\propto \phi^p$ near its minimum, 
one can show that, in equilibrium, the following virial relations between 
the kinetic, potential and the interaction energy densities hold (in this 
context, see, for example, Refs.~\cite{Lozanov:2016hid,Lozanov:2017hjm,
Maity:2018qhi}):
\begin{subequations}
\label{eq:virial}
\begin{eqnarray}
\f{1}{2}\,\l\langle\dot{\phi}^2 \r\rangle 
&=& \f{1}{2}\,\l\langle\frac{\vert\bm{\nabla}\phi\vert^2}{a^2}\r\rangle 
+ \f{p}{2}\, \l\langle V(\phi)\right\rangle
+ \l\langle V_{\mathrm{I}}(\phi,\mathcal{F})\r\rangle,\nn\\
\label{eq:virial1}\\
\f{1}{2}\, \l\langle \dot{\mathcal{F}}^2 \r\rangle 
&=& \f{1}{2}\,\l\langle\f{\vert\bm{\nabla}\mathcal F\vert^2}{a^2}\r\rangle
+ \l\langle V_{\mathrm{I}}(\phi,\mathcal F)\r\rangle,
\label{eq:virial2}
\end{eqnarray}
\end{subequations}
where the angular brackets indicate that the quantities have been averaged 
over space as well as the period of oscillation of the inflaton.  
During this backreaction phase, one can define the instantaneous EOS
averaged over the spatial volume as
\be
w =\f{\tf{1}{2}\, \dot{\phi}^2+\tf{1}{2}\,\dot{\mathcal{F}}^2 
-\tf{1}{6\,a^2}\,\vert\bm{\nabla} \phi\vert^2 
- \tf{1}{6\,a^2}\,\vert{\bm \nabla} \mathcal{F}\vert^2
- V_{\mathrm{I}}(\phi,\mathcal F)}{\tf{1}{2}\dot{\phi}^2 
+ \tf{1}{2}\,\dot{\mathcal F}^2 
+ \tf{1}{2\,a^2}\,\vert\bm{\nabla} \phi\vert^2 
+ \tf{1}{2\,a^2}\,\vert\nabla \mathcal F\vert^2 
+ V_{\mathrm{I}}(\phi,\mathcal F)}.\label{eq:eos_vir}
\ee
Upon using the virial relations~\eqref{eq:virial}, we find that the 
above expression for $w$ reduces to
\be
w = \f{1}{3} + \l(\f{p-4}{6}\r)\, \l(\f{p+2}{4} 
+ \f{\langle \rho_{\mathrm{G}}\rangle}{\langle V(\phi) \rangle} 
+\f{3\,\langle V_{\mathrm{I}}(\phi,\mathcal{F}\rangle}{2\,\langle 
V(\phi) \rangle}\r)^{-1},\label{eq:v-eos}
\ee
where $\langle \rho_{\mathrm{G}}\rangle = \langle \vert\bm{\nabla} 
\phi\vert^2/(2\,a^2)\rangle + \langle \vert\bm{\nabla} 
\mathcal{F}\vert^2/(2\,a^2)\rangle$ is the total energy density
associated with the spatial gradients in the fields.

\par

\begin{figure*}
\includegraphics[width=8.85cm]{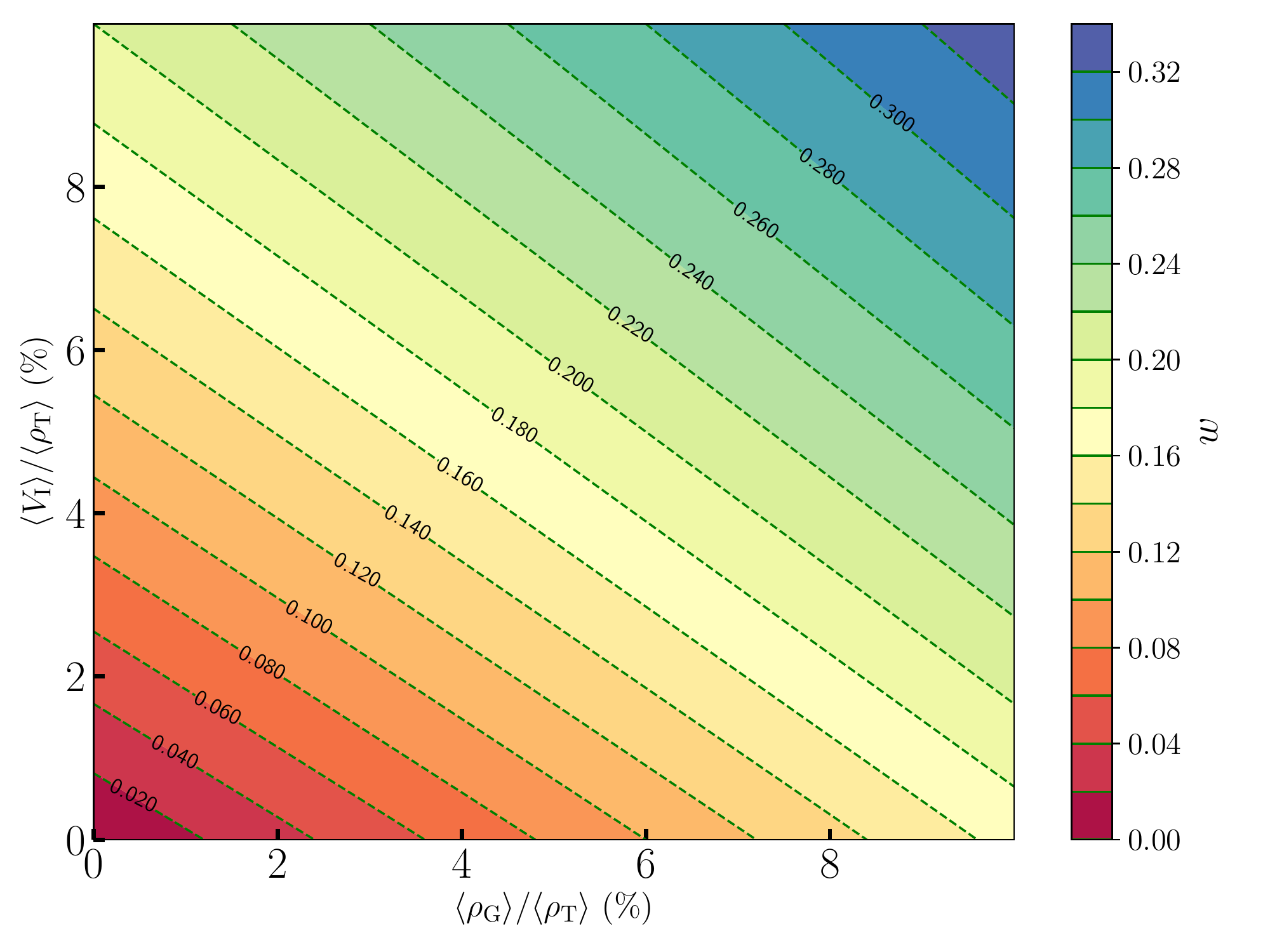}
\includegraphics[width=8.85cm]{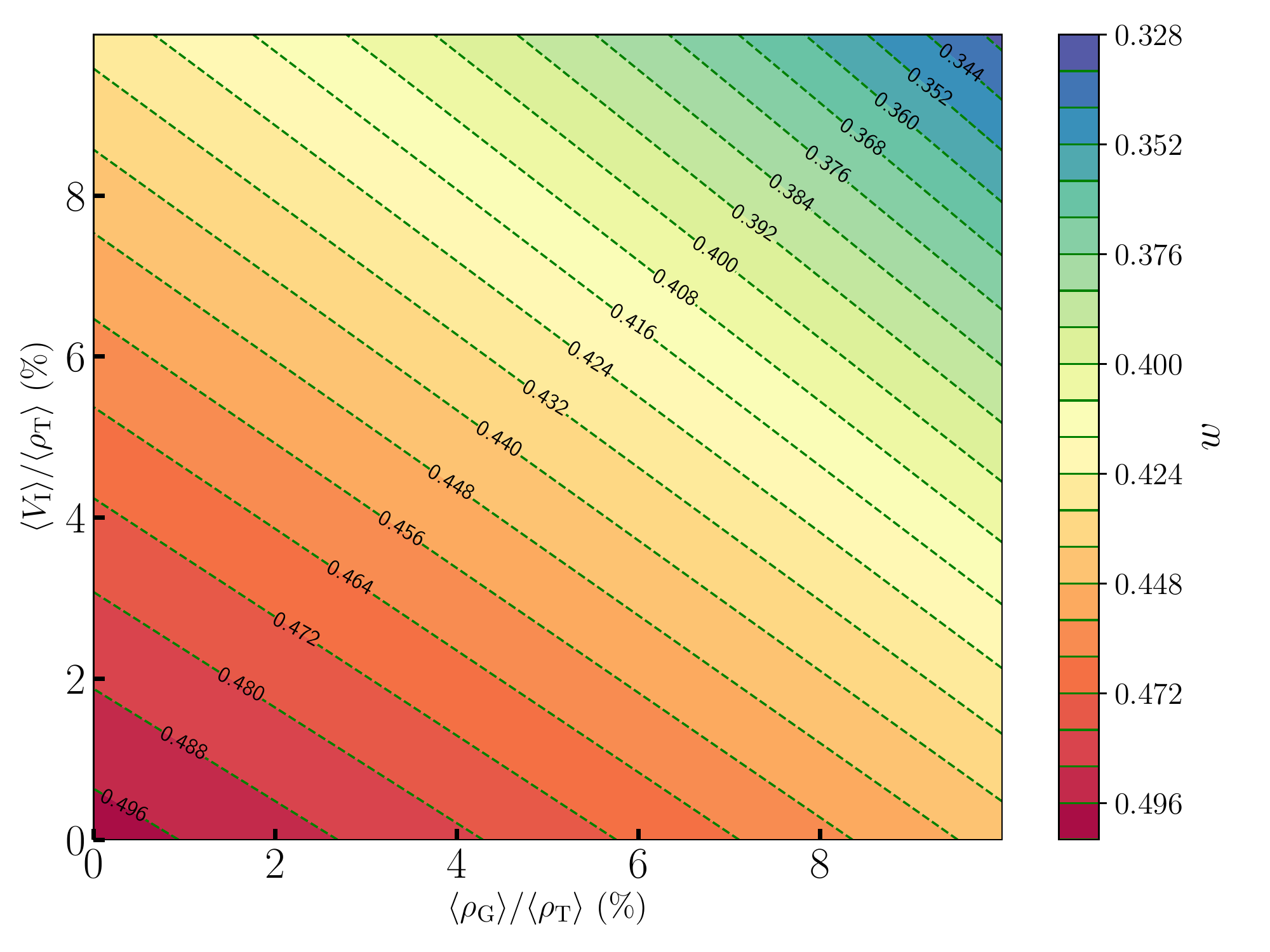}
\caption{The contours corresponding to constant EoS parameter~$w$ during
reheating--as defined by Eq.~\eqref{eq:v-eos}--have been plotted in 
the $\langle\rho_{\mathrm{G}}\rangle$--$\langle V_{\mathrm{I}}\rangle$ 
plane for the case of potentials $V(\phi)$ which behave as $\phi^2$ 
(on the left) and $\phi^6$ (on the right) near their minima.  
In order to make the axes dimensionless, we have divided both the axes by 
the quantity $\langle \rho_{\mathrm{T}}\rangle$, which represents the total 
energy density of the inflaton as well as the daughter fields. 
In plotting these contours, based on the results from lattice simulations, 
we have assumed that the kinetic energy density constitutes~$70\%$ of the
total energy density $\langle \rho_{\mathrm{T}}\rangle$~\cite{Maity:2018qhi}. 
However, we should hasten to add that changing the level of contribution due 
to the kinetic energy density does not alter the qualitative nature of the
plots.
As we have discussed, the transfer of energy to the daughter fields as well
as the growth of inhomogeneities occur rapidly at the end of the phase of 
coherent oscillations~\cite{Lozanov:2016hid,Lozanov:2017hjm}.  
Note that the plots clearly indicate that, as the contributions due to the 
gradient and the interaction energy densities increase, $w$ moves away from 
zero and eventually approaches~$1/3$.}\label{fig:v-eos}
\end{figure*}

It should be clear from the above equation for~$w$ that, as the gradient 
and the interaction energies begin to dominate, the second term in the 
expression becomes insignificant and the EOS parameter approaches~$1/3$.
To explicitly demonstrate these effects of the gradient and interaction energy 
densities on the EOS parameter, in Fig.~\ref{fig:v-eos}, we have plotted the 
contours of fixed~$w$ from Eq.~(\ref{eq:v-eos}) for potentials $V(\phi)$ which 
behave as $\phi^2$ and $\phi^6$ around their minima.  
There are two points that should be evident from the figure.
First, even a slight increase in the gradient or interaction energy densities
results in a nonzero instantaneous EOS parameter. 
Second, as we pointed out above, $w$ asymptotically approaches~$1/3$, as 
both the gradient and interaction energy densities increase. 
As we shall illustrate in the following subsection, these
expectations are corroborated by lattice simulations which allows one to 
track the EOS from the end of the phase of coherent oscillations to the 
beginning of the radiation dominated epoch (in this context, also see, 
for instance, Refs.~\cite{Lozanov:2016hid,Lozanov:2017hjm,Maity:2018qhi}).
These simulations suggest that, for $p<4$, the EOS parameter monotonically 
increases towards the asymptotic value of $1/3$.
Similarly, for $p > 4$, one finds that it decreases monotonically towards~$1/3$. 
From these arguments, we conclude that, in any realistic scenario, the EOS 
during reheating must be different than its value during the phase of coherent 
oscillations and that a vanishing EOS parameter is highly unlikely during 
this stage.


\subsection{Specific examples of the time-dependent EOS}\label{sec:se}

\begin{figure}
\begin{center}
\includegraphics[width=8.50cm]{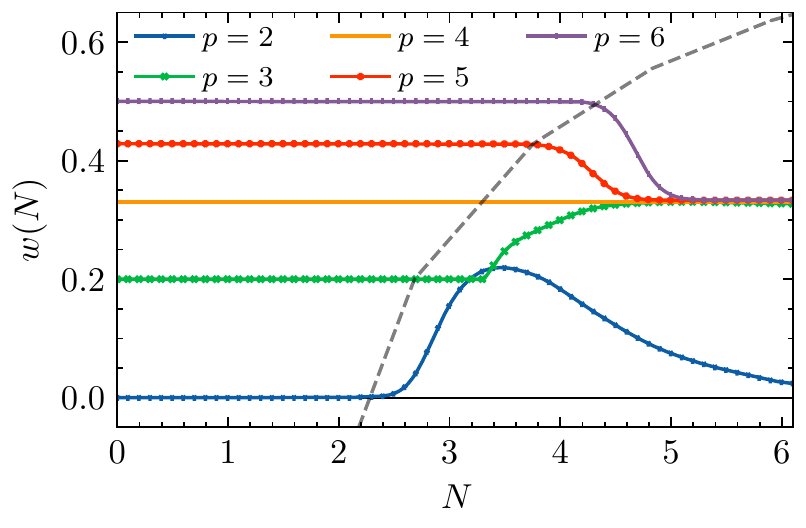}
\end{center}    
\vskip -15pt
\caption{The EOS parameter obtained from lattice simulations for the 
alpha-attractor T models described by the potential~(\ref{eq:tm}) has 
been plotted as a function of \e-folds for the cases wherein $p=(2,3,4,5,6)$. 
Notice that the EOS parameter starts with the value of $w_\mathrm{co}=(p-2)/(p+2)$
and approaches $w=1/3$ in all the cases apart from the case of $p=2$.
In the case of $p=2$, it is known that the specific coupling we have considered
is not effective to achieve radiation domination~\cite{Maity:2019ltu}. 
Utilizing Eq.~(\ref{eq:n_co_est}), in the figure, we have 
also plotted $w_\mathrm{co}=(p-2)/(p+2)$ against $N_{\mathrm{co}}$ (as a 
dashed black curve) for the case of $R_{\mathrm{co}}=25$.
The locations of the intersection of the curve with the EOS parameter indicate
the onset of the backreaction phase for the different indices~$p$.}
\label{fig:eos_lattice}
\end{figure}

In order to understand the typical form of the evolution of the EOS
parameter, in this subsection, we shall present the results of 
lattice simulations for a specific model.
We shall work with the so-called $\alpha$-attractor T models whose 
potentials are given by~\cite{Kallosh:2013hoa}
\begin{equation}
V(\phi) 
= \Lambda^4\, \l\vert\tanh^{p}\l(\f{\phi}{\sqrt{6\,\alpha}\,\Mpl}\r)\r\vert.
\label{eq:tm}
\end{equation}
We shall assume that the inflaton decays into a lighter scalar degree of 
freedom via a coupling of the form~$(g^2/2)\,\phi^2\,\chi^2$.
For the values of the parameter~$\alpha$ such that $\alpha \gtrsim 1$, the 
preheating dynamics in the above models are identical to that of the power 
law chaotic potentials $V(\phi) \propto \vert\phi\vert^p$ (in this context, 
also see Refs.~\cite{Lozanov:2016hid,Lozanov:2017hjm,Maity:2019ltu}).
With the above forms of the potential and interaction, we have solved for 
the coupled scalar field dynamics on a $256^3$ lattice using the parallel 
version of the lattice simulation 
code~\texttt{LATTICEEASY}~\cite{Felder:2000hq,Felder:2008zz}.
We have set $\alpha=1$ and $g^2=3.5\times10^{-7}$, and have considered the 
following five different values for the index describing the above
potential:~$p=(2,3,4,5,6)$. 
In these runs, the initial conditions for the background are set at the 
instant when the inflaton begins to oscillate near the bottom of the 
potential. 
The EOS parameter for the above two-field system is obtained using
Eq.~(\ref{eq:eos_vir}). 
In Fig.~\ref{fig:eos_lattice}, we have plotted the variation of the 
oscillation averaged EOS parameter against the number of \e-folds from 
the time when we start the simulations. 
As expected, we observe that, during the initial coherent oscillation phase,
the EOS parameter is given by $w_\mathrm{co}=(p-2)/(p+2)$. 
The EOS parameter begins to change once the inhomogeneities start to grow and 
it gradually tends towards $w=1/3$ in all the cases except for~$p=2$. 
We should point out that similar results from lattice simulations have also 
been arrived at earlier (see, for instance, Ref.~\cite{Antusch:2020iyq}).
For the case of $p=2$, in the scenario involving two fields, one finds that 
the final stage ends up being dominated by the inflaton itself, which 
restricts the the EOS parameter from approaching radiationlike behavior. 
Similar behavior has also been encountered when one takes into account only 
the self-resonance of the inflaton~\cite{Lozanov:2016hid,Lozanov:2017hjm}. 
However, the result for the case of $p=2$ must be viewed as a limitation of 
the specific coupling and, as the results for the other cases of~$p$ suggest, 
reheating can be expected to bring about a radiation dominated Universe. 
We must note that, in any realistic situation, reheating must lead to a 
radiation dominated phase, else one has to invoke an additional mechanism 
such as perturbative reheating to come to our aid.
The main goal of these lattice simulations is to motivate the construction of
time-dependent EOS parameter to describe the epoch of reheating.
In the following subsection, we shall discuss time-dependent EOS parameters
which effectively capture the results we have obtained from the end of the 
coherent oscillation phase until the beginning of the epoch of radiation 
domination.


\subsection{Parametrizing the EOS}\label{sec:param_eos}

Motivated by the results from the lattice simulations, in
order to capture the continuous variation of the EOS parameter from the end 
of coherent oscillations to radiation domination, we shall now parametrize 
the instantaneous EOS parameter {\it by hand}\/ in terms of \e-folds.
In choosing the functional form of the EOS parameter, we assume that it 
varies monotonically from its initial value~$w_{\mathrm{co}}$ to the final 
value of~$1/3$.
We find that this condition considerably restricts the form of the functions 
we can consider.

\par

We consider two different parametrizations of the following forms:
\begin{itemize}
\item 
Case A:~exponential form
\be
w(N) = w_0+ w_1\,\mathrm{exp}\l(-\f{1}{\Delta}\,\f{N}{N_{\mathrm{re}}}\right), 
\label{eq:eos-exp}
\ee 
\item 
Case B:~tan-hyperbolic form
\be
w(N) = w_0+ w_1\,\mathrm{tanh}\l(\frac{1}{\Delta}\,\frac{N}{N_{\mathrm{re}}}\r),
\label{eq:eos-tanh}
\ee
\end{itemize} 
where~$N$ is the number of \e-folds {\it counted from the end of the phase 
of coherent oscillations}.\/ 
We must clarify here that although, we have used the same 
symbols $w_0$, $w_1$ and $\Delta$ in the two parametrizations, {\it a priori}, we 
do not expect that two sets of parameters are related or that the functional 
dependence of $w(N)$ on them are similar.
The parameters $w_0$ and $w_1$ are fixed from the values of $w$ at the end of 
the coherent oscillations and the asymptotic limit which we take to be that 
of the radiation dominated epoch. 
Evidently, the parameter $\Delta$ controls the efficiency of the reheating 
process and determines how quickly the radiation dominated phase is attained. 
We further assume that the EOS parameter at the end of reheating, say,
$w_{\mathrm{re}}$, is within $10\%$ of the asymptotic value of~$1/3$.
There are two reasons for this assumption. 
The first is the reason that one has to account for various physical effects 
that can result in the deviation of the EOS parameter from~$1/3$ during the 
initial stages of radiation domination (see Ref.~\cite{Seto:2003kc}; in 
this context, also see Ref.~\cite{Weinberg:1972kfs}, Sec.~2.11). 
The second is the practical reason to set a benchmark where the energy density 
of radiation has formally begun to dominate the rest of the energy densities. 
We find that this choice of $w_{\mathrm{re}}$ fixes the value of~$\Delta$.
Under these conditions, the two parametrizations take the following form: 
\begin{equation}
w(N,p) 
= \begin{cases}
\f{1}{3} +\f{2}{3}\,\l(\f{p-4}{p+2}\r)\,
\mathrm{exp}\l(-\frac{1}{\Delta}\,\f{N}{N_{\mathrm{re}}}\r),
&\text{(A)}\\
\f{p-2}{p+2}-\f{2}{3}\,\l(\f{p-4}{p+2}\r)\,\mathrm{tanh}\l(\f{1}{\Delta}\,
\f{N}{N_{\mathrm{re}}}\r), 
&\text{(B)} 
\end{cases}\label{eq:eos-p}
\end{equation}  
with 
\begin{equation}
\f{1}{ \Delta} 
= \begin{cases}
\mathrm{ln}\l[\l(\f{p-4}{p+2}\r)\,
\l(\f{2}{3\,w_{\mathrm{re}}-1}\r)\r], & \text{(A)}\\
\tanh^{-1}\l\{\f{3}{2}\,\l[\f{p-2-w_{\mathrm{re}}(p+2)}{p-4}\r]\r\}. 
&\text{(B)}
\end{cases}\label{eq:Delta}
\end{equation}

\par

\begin{figure}
\includegraphics[width=8.50cm]{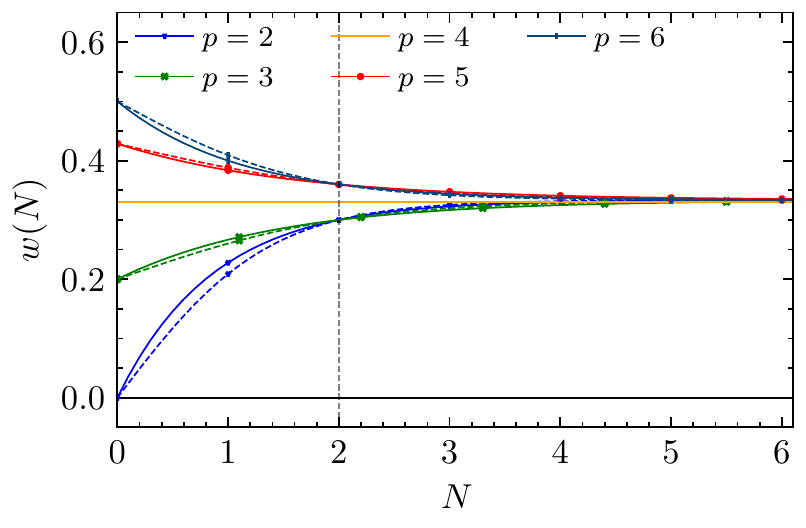}
\caption{The time-dependent EOS parameter $w(N)$ during reheating has been
plotted as a function of \e-folds when the inflaton evolves in potentials which
behave as $V(\phi)\propto\phi^p$ near their minima.
We have plotted both the exponential (as solid lines) and the tan-hyperbolic 
(as dashed lines) parametrizations that we have proposed 
[\cf Eq.~\eqref{eq:eos-p}] for a few different values of~$p$.
As a benchmark, we take the end of reheating (denoted by the red vertical line) 
to be when the EOS parameter reaches within $10\%$ of its asymptotic value 
of~$1/3$.}
\label{fig:wNp}
\end{figure}

We had already pointed out that, from its initial value of 
$w_\mathrm{co}=(p-2)/(p+2)$, the EoS parameter $w$ increases 
or decreases monotonically towards $1/3$ for $p<4$ and $p>4$,
respectively. 
For $p=4$, the reheating phase is indistinguishable from the 
radiation dominated epoch since $w_\mathrm{co}=1/3$.
Hence, in such a case, $\Delta\to 0$.
In Fig.~\ref{fig:wNp}, we have compared the two parametrizations 
described by Eq.~\eqref{eq:eos-p} for different values of~$p$. 
Note that the behavior of two parametrizations are qualitatively similar 
and, importantly, they broadly mimic the behavior we had seen in the results
from lattice simulations we had discussed in the previous subsection.

\par

Given the forms~\eqref{eq:eos-p} for the time varying EOS parameter, we can 
determine the corresponding $w_{\mathrm{eff}}$ [\cf Eq.~\eqref{eq:weff}] 
for the two cases $A$ and $B$ to be
\begin{equation}
w_{\mathrm{eff}}(p) 
=\begin{cases}
\f{1}{3} - \f{2\, \Delta}{3}\,
\l(\f{p-4}{p+2}\r)\,\l(\mathrm{e}^{-1/\Delta} - 1\r),
& \text{(A)}\\
\l(\f{p-2}{p+2}\r) - \f{2\, \Delta}{3}\,\l(\f{p-4}{p+2}\r)\, 
\mathrm{log}\,\l[\mathrm{cosh}\,\l(1/\Delta\r)\r]. 
& \text{(B)}\\
\end{cases}
\label{eq:weff-p}
\end{equation}
Thus, for a given inflationary potential, $w_{\rm eff}$ is fixed. 
In Tab.~\ref{tab:w_comp}, we have compared the values of $w_\mathrm{eff}$
for the two parametrizations with the value of $w_\mathrm{co}$ for a
set of values of~$p$.
\begin{table}
\begin{ruledtabular}
\begin{tabular}{cccc}
$p$ &$w_{\mathrm{co}}$ 
& $w^{\mathrm{exp}}_{\mathrm{eff}}$ & $w^{\mathrm{tanh}}_{\mathrm{eff}}$\\
\hline
$1$ & $-1/3$ & $0.12$ & $0.09$ \\
$2$ &  $0$   & $0.20$ & $0.19$\\ 
$4$ & $1/3$  & $1/3$   & $1/3$\\
$6$ & $1/2$  & $0.41$ & $0.42$\\
$8$ & $3/5$  & $0.44$ & $0.45$\\
$p\to\infty$ & $1$ & $0.53$ & $0.56$
\end{tabular}
\end{ruledtabular}
\caption{Comparison between the EOS parameter during coherent 
oscillations~$w_\mathrm{co}$ and the effective EOS parameter~$w_\mathrm{eff}$ 
during reheating for the two parametrizations that we have proposed.
Clearly, $w_\mathrm{eff}$ is largely independent of the two
parametrizations we have considered.
Also, note that, in general, $w_\mathrm{eff}$ proves to be 
substantially different from $w_\mathrm{co}$.}
\label{tab:w_comp}
\end{table}
On substituting Eq. \eqref{eq:weff-p} for $w_\mathrm{eff}$ 
in Eq.~\eqref{eq:Nre} for~$N_\mathrm{re}$, we obtain that
\begin{equation}
N_{\rm re} = 
\begin{cases}
2\,\l(\f{p+2}{p-4}\r)\,\mathcal{N}
\l[\Delta\,\l(1- \mathrm{e}^{-1/\Delta}\r)\r]^{-1}, & \text{(A)}\\
2\,\l(\f{p+2}{p-4}\r)\,\mathcal{N}
\l\{1 -\Delta\log\,\l[\cosh\,(1/\Delta)\r]\r\}^{-1}, & \text{(B)}
\end{cases}\label{eq:Nre-f}
\end{equation}
where the quantity $\mathcal{N}$ is defined as
\begin{eqnarray}
\mathcal{N} 
&=& N_{k} + \f{(4-p)}{2\,(p+2)}\,N_{\mathrm{co}}
+\ln \l(\frac{k}{a_{0}\, T_{0}}\r)
+\f{1}{4}\ln\l(\f{30}{\pi^{2}\, g_{\mathrm{re}}}\r)\nn\\ 
& &\,+ \f{1}{3}\, \ln \l(\frac{11\, g_{{\mathrm{s},\mathrm{re}}}}{43}\r)
- \ln \l(\f{H_{k}}{\rho_{\mathrm{end}}^{1/4}}\r),\label{eq:calN}
\end{eqnarray}
while $\Delta$ is given by Eq.~\eqref{eq:Delta}.
It should be clear from the above equation that, barring $g_\mathrm{re}$,
$g_{\mathrm{s},\mathrm{re}}$ and $N_\mathrm{co}$, the duration of 
reheating~$N_\mathrm{re}$ depends only on the inflationary parameters and 
the CMB observables.
On substituting the above expressions for $w_\mathrm{eff}$ and $N_\mathrm{re}$
in Eq.~\eqref{eq:Tre2}, we can arrive at the corresponding reheating 
temperature~$T_\mathrm{re}$.


\section{Application to an inflationary model}\label{sec:im}

Let us now apply our arguments to an inflationary model which has the 
desired behavior near its minima.
Towards this end, we shall consider the so-called $\alpha$-attractor 
model described by potential~\cite{Carrasco:2015pla,Carrasco:2015rva}
\begin{equation}
V(\phi) 
= \Lambda^4\,
\l[1-\mathrm{exp}\,\l(-\sqrt{\frac{2}{3\,\alpha}}\,\f{\phi}{\Mpl}\r)\r]^p, 
\label{eq:V-aa}
\end{equation}
where $\Lambda$, $\alpha$ and $p$ are, evidently, parameters that 
characterize the model.
As we had pointed out, we can express the first slow roll parameter as
$\epsilon\simeq (\Mpl^2/2)\,(V_\phi/V)^2$, where the subscript $\phi$
denotes the derivative of the potential with respect to the field.
Let us define the second slow roll parameter as $\eta\simeq \Mpl^2\,
(V_{\phi\phi}/V)$.
Then, in the slow roll approximation, the inflationary observables---\viz 
the scalar spectral index~$n_{_{\mathrm{S}}}$ and the 
tensor-to-scalar ratio~$r$---can be expressed in terms of these 
parameters as (see, for instance, the reviews~\cite{Mukhanov:1990me,
Martin:2003bt,Martin:2004um,Bassett:2005xm,Sriramkumar:2009kg,
Sriramkumar:2012mik,Baumann:2009ds,Linde:2014nna,Martin:2015dha})
\begin{subequations}\label{eq:ip}
\begin{eqnarray}
n_{_{\mathrm{S}}} &=& 1 -6\,\epsilon_k + 2\,\eta_k, \label{eq:ns-sr}\\ 
r &=& 16\,\epsilon_k,~\label{eq:r-sr}
\end{eqnarray}
\end{subequations}
where the subscript $k$ indicates that these quantities have to be
evaluated when the mode leaves the Hubble radius.
Moreover, the scalar amplitude $A_{_{\mathrm{S}}}$ can be expressed 
in terms of the value of the Hubble parameter~$H_k$ and the 
tensor-to-scalar ratio~$r$ as follows:
\begin{equation}
H_k = \sqrt{\frac{r\,A_s}{2}}\,\pi\,\Mpl.\label{eq:Hk-As}
\end{equation}

\par

The number of \e-folds $N_k$ between the mode $k$ leaving the Hubble radius 
and the end of inflation can be expressed in the slow roll approximation as
\begin{equation}
N_k = \int_{\phi_k}^{\phi_{\mathrm{end}}}\d \phi\, \frac{H}{\dot\phi}
\simeq \f{1}{\Mpl^2}\int^{\phi_k}_{\phi_{\mathrm{end}}}\d\phi\,
\frac{V}{V_\phi}.
\end{equation}
For the inflationary potential~\eqref{eq:V-aa} of our interest, $N_k$ can be
evaluated to be
\begin{eqnarray}
N_k &= &\f{3\,\alpha}{2\,p}\,
\biggl[\mathrm{exp}\l(\sqrt{\f{2}{3\,\alpha}}\,\frac{\phi_k}{\Mpl}\r)
-\mathrm{exp}\l(\sqrt{\f{2}{3\,\alpha}}\,\f{\phi_{\mathrm{end}}}{\Mpl}\r)\nn\\ 
& &-\, \sqrt{\f{2}{3\,\alpha}}\,\frac{(\phi_k -\phi_{\mathrm{end}})}{\Mpl}\biggr].
\label{eq:Nk}
\end{eqnarray}
The quantity $\phi_\mathrm{end}$ can be determined by the condition $\epsilon = 1$
and is given by
\begin{equation}
\f{\phi_\mathrm{end}}{\Mpl} 
= \sqrt{\frac{3\,\alpha}{2}}\, \mathrm{ln}\l(1 + \f{p}{\sqrt{3\, \alpha}}\r)
\end{equation}
so that we have
\begin{equation}
V_{\mathrm{end}} = V(\phi_\mathrm{end})
=\Lambda^4\, \l(\f{p}{p+\sqrt{3\,\alpha}}\r).
\end{equation}
The above relations between~$\phi_k$, $\phi_\mathrm{end}$ and $N_k$ and Eq.~\eqref{eq:ns-sr} for the scalar spectral index allows us to write 
$n_{_{\mathrm{S}}}$ in terms of $N_k$.
We can then invert the relation to express $N_k$ in terms of~$n_{_{\mathrm{S}}}$.

\par

Note that, near its minimum, the inflationary potential~\eqref{eq:V-aa} 
can be approximated as
\be
V(\phi)\simeq \Lambda^4\,\left(\frac{2\,\phi}{3\,\alpha\,\Mpl}\r)^p,
\ee
which is of the form we desire.
We should mention here that, for $\alpha=1 $ and $p=2$, the
potential~\eqref{eq:V-aa} corresponds to the Starobinsky model, 
which is the most favored model according to the recent CMB
observations~\cite{Ade:2015lrj,Akrami:2018odb}.  
Recall that, for a given $p$, $w_\mathrm{eff}$ is fixed [cf. Eq.~\eqref{eq:weff-p}].
Hence, we have most of the required ingredients to calculate the 
duration of reheating~$N_\mathrm{re}$ and the corresponding reheating 
temperature~$T_\mathrm{re}$ using the expressions~\eqref{eq:Nre-f} 
and~\eqref{eq:Tre2}.
However, we shall require values for $g_\mathrm{re}$, $g_{\mathrm{s}, 
\mathrm{re}}$ and $N_\mathrm{co}$.
It seems reasonable to choose $g_\mathrm{re}=g_{\mathrm{s}, \mathrm{re}}
=10^2$~\cite{Mukhanov:2005sc}.

\par

Let us now turn to identifying a suitable choice for~$N_\mathrm{co}$. 
The duration of the phase of coherent oscillation can strongly depend 
on the model parameters and, importantly, on the couplings of the 
inflaton to other fields~\cite{Figueroa:2016wxr}. 
In particular, if nonperturbative processes dominate throughout the reheating 
phase, thermalization may be achieved within a few \\e-folds making it difficult 
to connect the reheating phase with the CMB observables.   
However, this phase can be inefficient or 
delayed~\cite{GarciaBellido:2008ab,Repond:2016sol,Freese:2017ace} and can 
result in the breakdown of coherent oscillations without 
thermalization~\cite{Easther:2010mr,Musoke:2019ima}. 
It has been pointed out that the final stage of reheating must
involve the perturbative decay of the inflaton and the thermalization is slow if 
we, for example, invoke certain supersymmetric extensions of the standard model 
to achieve inflation~\cite{Allahverdi:2005mz}. 
Although preheating can generate a plasma of inflaton and other daughter fields 
in kinetic equilibrium, complete thermal equilibrium is achieved over a much 
larger time scale than that of the preheating~\cite{Felder:2000hr,Micha:2004bv}. 
The exact number for $N_{\mathrm{co}}$ will depend on the inflationary potential 
as well as the type of interaction(s) and the strength of the coupling 
parameter(s). 
However, if there are no daughter fields present and inflaton fragments only 
due to self-resonance, one can arrive at an estimate for the upper bound on its 
value, which is found to be (see Refs.~\cite{Lozanov:2016hid,Lozanov:2017hjm})
\begin{equation}
N_{\mathrm{co}} \simeq \f{(p+2)}{6}\,\ln\,R_{\mathrm{co}},
\label{eq:n_co_est}
\end{equation}
where $R_{\mathrm{co}} \sim \mathcal{O}(10^2)$ depends on the resonance structure. 
On assuming $R_{\mathrm{co}}$ to be $10^2$, we get $N_{\mathrm{co}}=(3.07, 4.60,
6.14)$ for $p=(2,4, 6)$. 
If coupling to other fields are present, the value of $N_{\rm co}$ will naturally
decrease~\cite{Figueroa:2016wxr}. 
Therefore, the period of preheating is negligible compared to the entire duration 
of reheating.
Due to these reasons, we consider $N_\mathrm{co}$ to be small and set it to unity. 

\par

\begin{figure*}
\includegraphics[width=8.10cm]{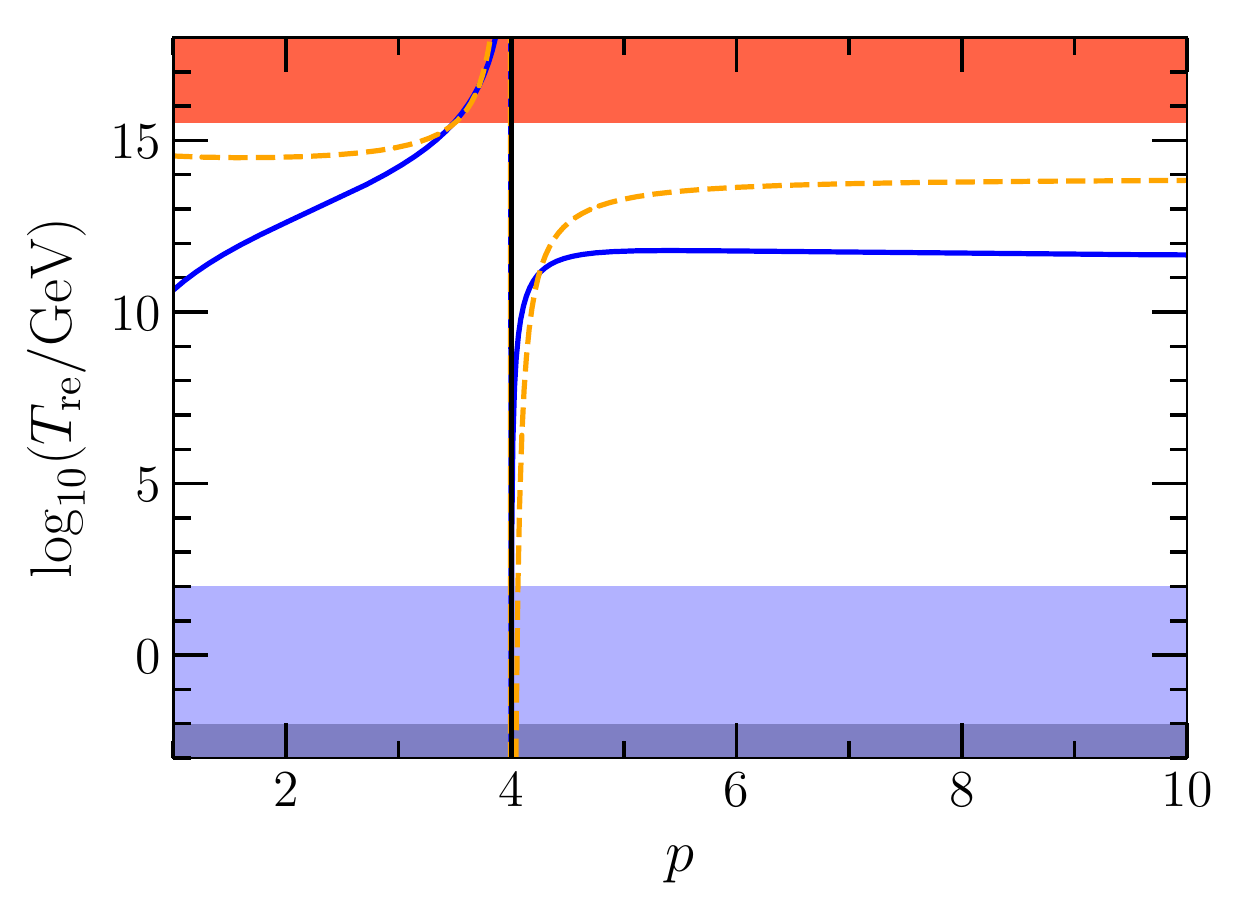}
\includegraphics[width=9.00cm]{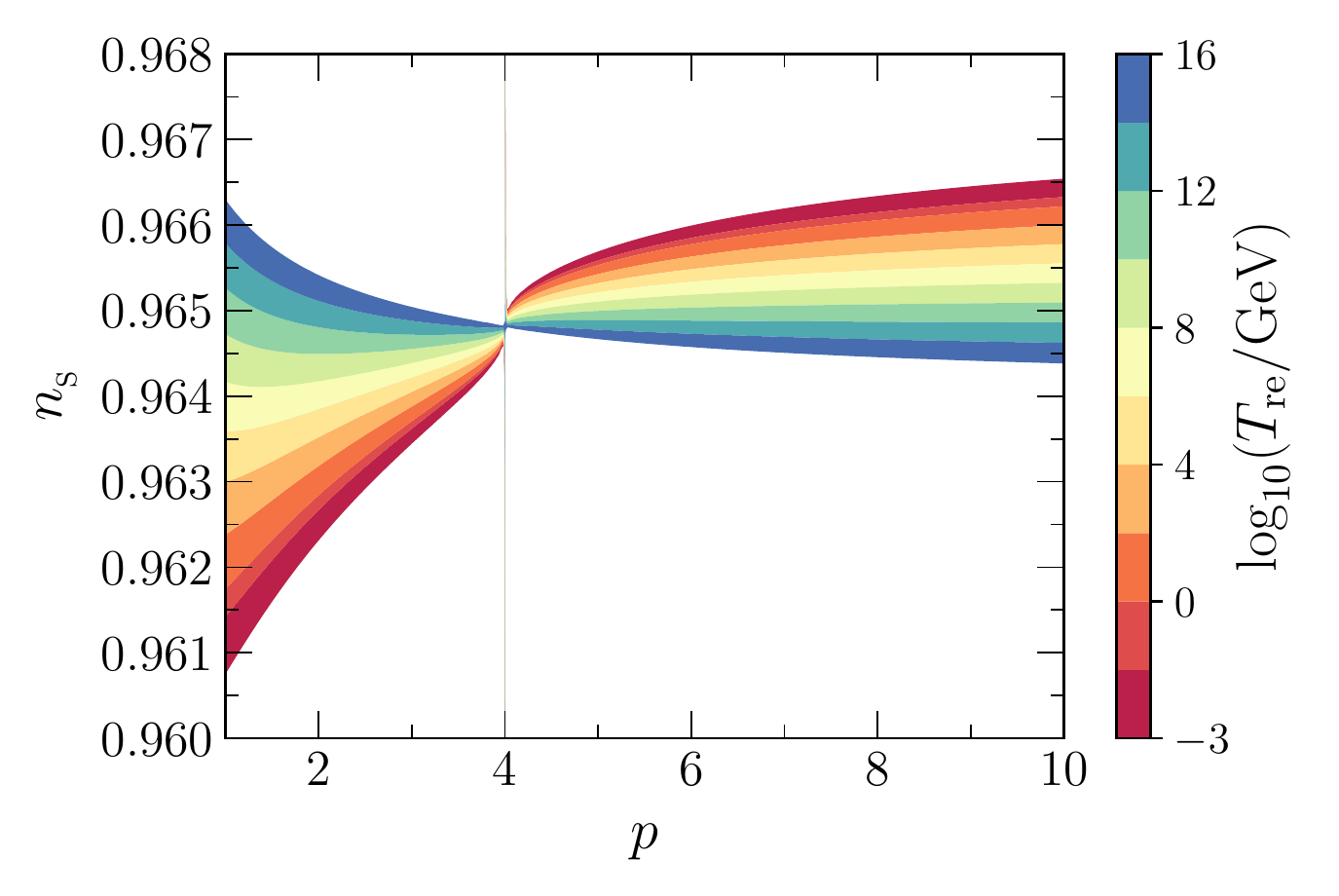}
\caption{The dependence of the reheating temperature $T_\mathrm{re}$ on the 
index~$p$ has been illustrated (on the left) for $n_{_{\mathrm{S}}}= 0.9649$
which leads to the best-fit to the CMB data~\cite{Ade:2015lrj,Akrami:2018odb}.
We have plotted the dependence of $T_\mathrm{re}$ on $p$ for $w_\mathrm{eff}$ 
corresponding to the exponential parametrization [cf. Eqs.~\eqref{eq:eos-p}
and~\eqref{eq:weff-p}] (as the solid blue curve) as well as for the choice 
$w_\mathrm{eff}=w_\mathrm{co}=(p-2)/(p+2)$ (as the dashed orange curve). 
We have also indicated the following domains (in the figure on the left):~the 
region above maximum possible reheating temperature of
$T_\mathrm{inst}=\l[30\,\rho_\mathrm{end}/(g_\mathrm{re}\,\pi^2)\r]^{1/4}$
corresponding instantaneous reheating or $N_\mathrm{re}= 0$ (in red), the
domains below the electroweak scale taken to be $T_{_{\mathrm{EW}}}\sim 100\, 
\mathrm{GeV}$ (in lighter blue) and the region below $10\, \mathrm{MeV}$ which 
is the minimum temperature required for BBN (in darker blue).
Moreover, we have illustrated the dependence of $T_\mathrm{re}$ on $n_{_{\mathrm{S}}}$
and $p$~(on the right) for the choice of $w_\mathrm{eff}$ corresponding to the
exponential parametrization. 
Note that we have set $\alpha = 1$ in both these plots.}\label{fig:Tre}
\end{figure*}

With all these necessary ingredients at hand, let us now compute the reheating
temperature $T_\mathrm{re}$ for the model of our interest.
Note that, $T_\mathrm{re}$ depends on $n_{_{\mathrm{S}}}$, $p$, $\alpha$, and 
$w_\mathrm{eff}$.
We shall set $\alpha=1$ without any loss of generality. 
Since $w_\mathrm{eff}$ is largely independent of the two parametrizations 
[cf. Tab.~\ref{tab:w_comp}], we shall choose to work with the values 
corresponding to the exponential form for~$w(N)$.
In Fig.~\ref{fig:Tre}, we have highlighted the dependence of $T_\mathrm{re}$ 
on~$n_{_{\mathrm{S}}}$ and~$p$ in two different manner.
We have first plotted $T_\mathrm{re}$ as a function of~$p$ for the value 
of~$n_{_{\mathrm{S}}}$ that leads to the best fit to the recent CMB 
data~\cite{Ade:2015lrj,Akrami:2018odb}.
In the figure, we have also illustrated the simultaneous dependence 
of~$T_\mathrm{re}$ on~$n_{_{\mathrm{S}}}$ and~$p$.
Note that the lower bound on the reheating temperature is arrived at from the 
BBN constraints as $T_\mathrm{BBN} \sim 10\,\mathrm{MeV}$ (see 
Refs.~\cite{Kawasaki:1999na,Steigman:2007xt,Fields:2014uja}; for a recent 
discussion, see Ref.~\cite{Hasegawa:2019jsa}), whereas the upper limit comes 
from the condition of instantaneous reheating which corresponds to the 
inflationary energy scale of the order of the grand unified theory (GUT) scale of about $10^{16}\, 
\mathrm{GeV}$ that arises in certain supersymmetric theories.

\par

Let us emphasize a few more points concerning Fig.~\ref{fig:Tre}.
It is clear that the new effective EOS parameter we have arrived at lowers 
the reheating temperature.
This effect can be attributed to the dependence of $T_\mathrm{re}$ on the 
ratio $(1+w_{\mathrm{eff}})/(3\,w_\mathrm{eff} -1)$, which is always higher 
than the one computed with $w_\mathrm{eff} = w_\mathrm{co}=(p-2)/(p+2)$ for 
a given value of $p$.
Thus, our proposal for the time-dependent EOS and its effect can, in principle, 
be tested in future experiments~\cite{Finelli:2016cyd,Kuroyanagi:2014qza}.
Moreover, note that, the variation of $T_\mathrm{re}$ with $p$ also depends 
on the value of the scalar spectral index. 
It is evident from Fig.~\ref{fig:Tre} that, for $p<4$, an increase in the 
value of $n_{_{\mathrm{S}}}$ results in a larger value of~$T_\mathrm{re}$. 
This is due to the fact that for $p<4$, $w_\mathrm{eff} < 1/3$ and, hence, 
an increase in the value of $n_{_{\mathrm S}}$ leads to a smaller value of
$N_\mathrm{re}$ which, in turn, leads to a larger value of $T_\mathrm{re}$. 
However, for $p>4$, the conditions are reversed and we have a decreasing 
$T_\mathrm{re}$ for an increasing $n_{_{\mathrm S}}$. 


\section{Discussion}\label{sec:conc}

In this work, we have computed the effective EOS parameter during 
the reheating phase of the Universe by taking into account the 
time evolution of the instantaneous EOS parameter. 
We have shown that the gradient and interaction energy densities force the
instantaneous EOS parameter to deviate from its value during the phase of
coherent oscillations which succeeds inflation. 
Assuming that the inflationary potential behaves as $V(\phi)\propto \phi^p$ 
about its minimum, based on results from lattice simulations,
we have argued that, during reheating, $w$ increases monotonically and
approaches~$1/3$ for $p<4$, whereas, for $p>4$, it decreases monotonically 
to~$1/3$ (cf. Fig.~\ref{fig:eos_lattice}). 
In order to capture such a behavior, we have proposed two different functional 
forms of the time varying EOS parameter during reheating
(cf. Fig.~\ref{fig:wNp}). 
We find that the resulting value of $w_\mathrm{eff}$ depends only on the
inflationary model parameter~$p$ and is largely independent of the
parametrizations we have considered for~$w(N)$~(cf. Tab.~\ref{tab:w_comp}).

\par

Let us stress here a few further points concerning the results we have obtained.
Note that, in our approach, $w_\mathrm{eff}$ is {\it completely}\/ determined by 
the inflationary parameter~$p$. 
Therefore, for a specific $p$, the reheating temperature $T_\mathrm{re}$ is fixed
for a given value of the scalar spectral index~$n_{_{\mathrm S}}$.
This should be contrasted with earlier studies, wherein there is an arbitrariness
in choosing the value of~$w_\mathrm{eff}$. 
As we discussed earlier, often $w_{\rm eff}$ is either assumed to lie in the 
range $-1/3 \leq w_\mathrm{eff} \leq 1$ or simply taken to be same as that of 
$w_\mathrm{co}$. 
However, various (p)reheating studies have indicated towards time varying EOS,
which has been captured efficiently with our parametrization. 
With such a time varying EOS parameter, we can uniquely define $w_\mathrm{eff}$ 
which, as we highlighted, is fixed by the behavior of the field around the 
minimum of the potential.
It is worth stressing again that the $w_\mathrm{eff}$ we have arrived at is 
largely independent of parametrization. 
Thus, this study mitigates the arbitrariness in defining the effective EOS 
parameter during reheating for a given inflationary model. 

\par
 
Lastly, note that, though we have worked with the $\alpha$-attractor model of
inflation specified by the potential~\eqref{eq:V-aa}, our analysis applies to
all the inflationary models which behave as $\phi^p$ around their minima. 
With ongoing and forthcoming CMB missions expected to constrain the inflationary
parameters more accurately, we believe that our proposal for the time-dependent
EOS during reheating can be well tested in the near future.


\section*{Acknowledgements}

The authors thank Debaprasad Maity for fruitful discussions. 
P. S. wishes to thank the Indian Institute of Technology Madras, Chennai, India, 
for support through the Institute Postdoctoral Fellowship.
S. A. is supported by the National Postdoctoral Fellowship of the Science and
Engineering Research Board (SERB), Department of Science and Technology (DST), 
Government of India (GOI). 
L. S. wishes to acknowledge support from SERB, DST, GOI, through the Core Research
Grant No. CRG/2018/002200.

\appendix

\section{Behavior of the EOS parameter during preheating}
\label{appendix:appA}

\begin{figure}[!t]
\begin{center}
\includegraphics[width=8.50cm]{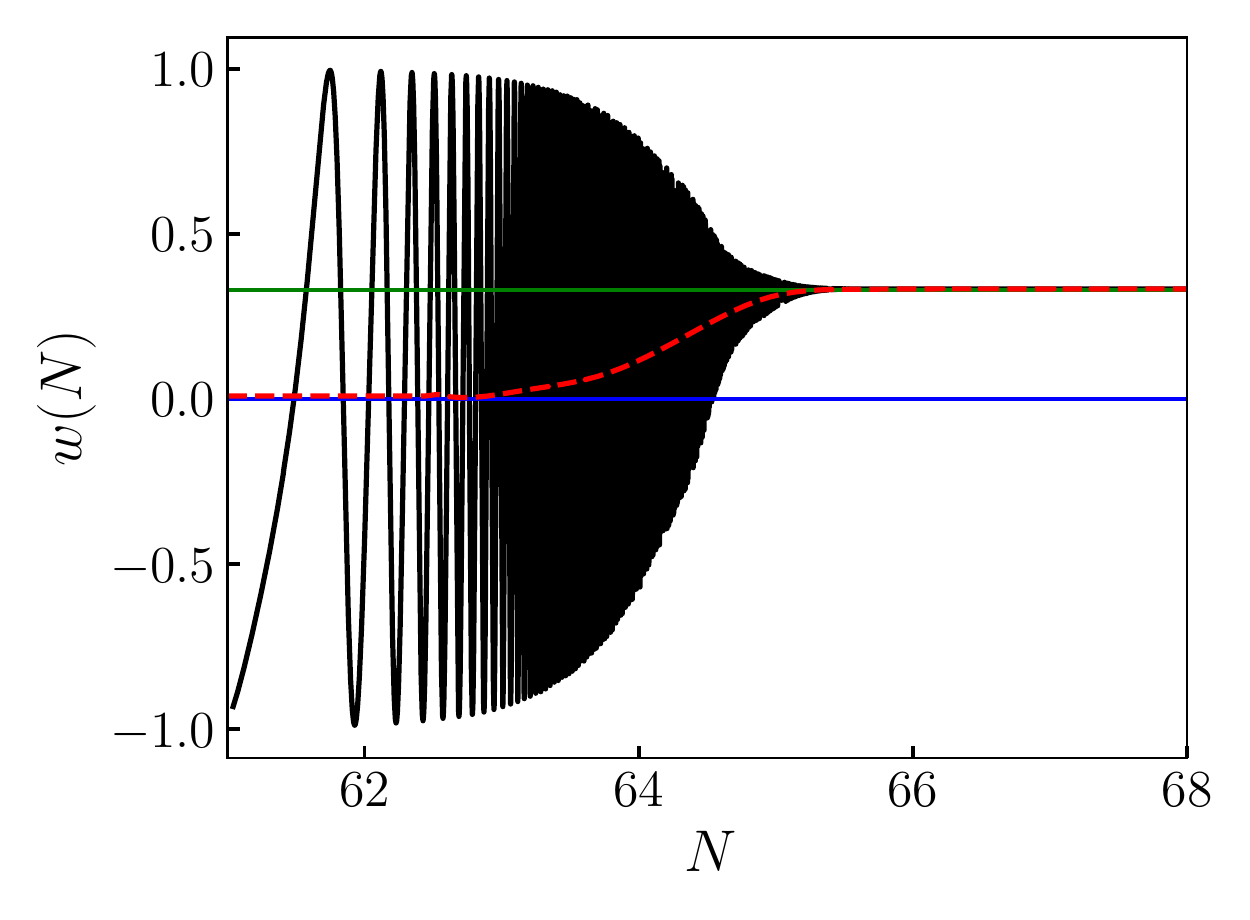}
\end{center}
\vskip -15pt
\caption{The behavior of the instantaneous EOS parameter (in black) as well as
the corresponding quantity arrived at after averaging over the oscillations (in 
orange) have been plotted for the case of the popular quadratic inflationary 
potential.
We have set the value of the decay width to be $\Gm=2\times10^{-9}\,\Mpl$ in 
arriving at these plots.
Note that the averaged EOS parameter starts at $w=0$ (indicated by the blue 
horizontal line) and eventually approaches $1/3$ (indicated by the green 
horizontal line) suggesting the eventual transfer of energy from the inflation 
to radiation.}
\label{fig:eos-p-ph}
\end{figure}

In order to highlight the behavior of the EOS parameter during the epoch of
coherent oscillations that immediately follows inflation, in this Appendix, 
we shall briefly discuss the well known case of perturbative reheating in 
the popular quadratic inflationary potential (in this context, see also
Refs.~\cite{Mukhanov:2005sc,Podolsky:2005bw,Martin:2010kz}). 
Incidentally, the perturbative mechanism we shall discuss here is the original 
idea of reheating, as we had mentioned in the introduction~\cite{Albrecht:1982mp}. 
In this case, to transfer the energy from the inflaton to radiation, an additional
damping term is introduced by hand in the equation governing the inflaton in the 
following fashion:
\begin{equation}
\ddot{\phi} + 3\,H\,\dot{\phi} + \Gm\,\dot{\phi} + V_\phi = 0,\label{eq:inf_decay}
\end{equation}
Physically, the damping term is expected to arise due to quantum particle creation 
as the inflaton decays into other lighter species that are coupled to it. 
For instance, if $\phi$ is allowed to decay into fermionic channels via an 
Yukawa interaction of the form $\mathcal{L}_{\mathrm{int}} \supset
g\,\phi\,\bar{\psi}\,\psi$, then, using the methods of perturbative quantum field
theory, one can show that~\cite{Schwartz:2013pla}
\begin{equation}
\Gm\equiv\Gamma_{\phi\to\bar{\psi}\,\psi}=\f{g^2\,m_{\phi}}{8\,\pi},\label{eq:p2ss}
\end{equation}
where $m_{\phi}$ is the tree-level mass of the inflaton.
Since we require a radiation dominated Universe after reheating, for simplicity,
it is often assumed that the inflaton directly transfers it's energy to radiation.
In order to conserve the total energy of the systems involved, as a consequence 
of the additional decay term in the equation of motion~(\ref{eq:inf_decay}) of 
the inflaton, the equation describing the conservation of energy density~$\rho_\gamma$
of radiation is modified to be
\begin{equation}
\dot{\rho_\gamma} + 4\,H\,\rho_{\gamma} - \Gm\, \dot{\phi}^2 = 0,
\label{eq:rad_den}
\end{equation}
while the Hubble parameter $H$ is governed by the following Friedmann equation:
\begin{equation}
H^2 = \f{1}{3\,\Mpl^2}\,\l(\f{\dot{\phi}^2}{2} + V(\phi) + \rho_\gamma\r).
\label{eq:Hps}
\end{equation}
The system of Eqs~\ref{eq:inf_decay}) and (\ref{eq:rad_den}) can be readily
solved with the initial conditions on the inflaton imposed at end of inflation and
the initial radiation density assumed to be zero. 
The exact EOS parameter for the system is defined as
\begin{equation}
w = \frac{\tf{1}{2}\,\dot{\phi}^2 - V(\phi)+\tf{1}{3}\rho_{\gamma}}{\tf{1}{2}\,\dot{\phi}^2
+ V(\phi)+\rho_\gamma\hfill}.\label{eq:inst_eos}
\end{equation}
We have plotted the EOS parameter during preheating for the case of the quadratic
potential in Fig.~\ref{fig:eos-p-ph}.
It is clear that the additional coupling introduced by hand transfers the energy 
from the inflation to radiation fairly effectively, in fact within a matter of a
few \e-folds.
\clearpage
%
\end{document}